\documentclass{aa}
\usepackage[varg]{txfonts}
\usepackage{graphicx}
\usepackage{amsmath,amssymb,pgf,graphicx,color,url}
\usepackage[colorlinks=true,citecolor=blue,linkcolor=blue]{hyperref}

\usepackage{natbib}
\bibpunct{(}{)}{;}{a}{}{,} 

\begin{document}
\title{The anti-glitching gamma-ray pulsar PSR J1522-5735}
\author{A.G.~Panin \thanks{E-mail: \email{panin@ms2.inr.ac.ru}}
\and E.V.~Sokolova \thanks{E-mail: \email{sokol@ms2.inr.ac.ru}}}
\institute{Institute for Nuclear Research of the Russian Academy of Sciences,
  Moscow 117312, Russia\label{inst1}}
\date{Received <date> / Accepted <date>}

\abstract
    {
      A small number of pulsar glitches have been identified as
      anti-glitches or spin-down glitches, where the overall contribution to
      the pulsar’s rotation frequency is negative. A notable example of a
      spin-down glitch was observed in the rotation-powered pulsar PSR
        J1522-5735, a radio-quiet gamma-ray
      pulsar discovered by blind searches in the three-year data from the
      {\it Fermi} Large Area Telescope (LAT).
    }
    {
      This work aims to search for PSR J1522-5735's glitches using
      {\it Fermi}-LAT data from over 15 years of observations.
    }
    {
      The weighted H-test statistic was applied to identify glitches and
      evaluate the related changes in pulsar's spin parameters. The timing
      solution based on these results was further refined by maximisation of
      the unbinned likelihood. The Bayesian information criterion was used to
      set an appropriate number of parameters in the timing solution to avoid
      overfitting.
      
    }
    {
      The analysis revealed eight glitch events: a regular spin-up glitch,
      a spin-up glitch over-recovered to a spin-down, and six anti-glitches.
      These events were radiatively quiet, exhibiting no significant
      variations in the shape of the pulse profile or energy flux.
    }
    {
      The results may suggest that an internal mechanism is responsible for
      spin-down glitch phenomena.
    }

\maketitle

\section{Introduction}

Pulsars are fast rotating neutron stars. Dipole radiation and particle winds
cause pulsars to slow down at an extremely steady rate, making them amongst
the most precise means of measuring time in the Universe. Nevertheless, many
pulsars exhibit sudden step-like increases of rotation frequency known as
glitches~\citep{Manchester, Downs, Espinoza}. There are two main models
to explain glitches. In the starquake glitch model of~\citet{Baym},
an oblate pulsar crust deforms towards an almost  spherical shape as
the pulsar slows down. This leads to a sudden crack in the crust and a
decrease in the moment of inertia, resulting in a sudden increase in
the angular velocity of the pulsar. In the
model proposed by~\citet{Anderson} \citep[see also][]{Pines, Link},
glitches are associated with angular momentum transfer
from the faster rotating superfluid stellar interior to the neutron star
crust.

Recent timing observations of the magnetar 1E~2259+586 by~\cite{Archibald}
\citep[see also][]{Icdem} revealed a sudden spin-down event (i.e. anti-glitch).
The anti-glitch was accompanied by a hard X-ray burst, which was detected by the
{\it Fermi} Gamma-ray Burst Monitor~\citep{Foley}, with a subsequent
increase in the flux and a moderate change in the pulse
profile~\citep{Archibald}. However, subsequent monitoring of the magnetar
1E~2259+586 led to the detection of another anti-glitch \citep{Younes} with
no evidence for flux enhancement or change in the spectral or pulse profile
shape.

To date, eight anti-glitches have been reported across three magnetars,
one accretion-powered pulsar, and one rotation-powered pulsar. 
The anti-glitched magnetars are 1E 2259+586, 1E 1841-045, and XTE J1810-197.
In the magnetar 1E~1841-045, a sudden spin-down occurred with no significant
variations in the pulsed X-ray output~\citep{SasmazMus}. The anti-glitch
in the transient magnetar XTE J1810-197 was accompanied by alterations in
the pulse profile with no radiative changes in its emission~\citep{Pintore,
  Vurgun}. The ultraluminous accreting X-ray pulsar NGC 300 ULX-1 exhibited
three radiation-quiet anti-glitches~\citep{Ray}. A permanent spin frequency
decrease in the rotation-powered pulsar PSR B0540-69 was identified without
any major alteration in the pulse profile or any significant increase of
the pulsed flux~\citep{Tuo}.  

Undoubtedly, the unexpected occurrence of anti-glitches poses a significant
challenge to standard glitch theories. The close correlation observed between
some anti-glitches and outburst activity suggest that these events may be
due to the influence of external processes, such as strong
outflows~\citep{Tong}, a sudden twisting of the magnetic field
lines~\citep{Lyutikov}, or the accretion of orbiting
objects~\citep{Katz, Huang}. In contrast, the identification of
radiation-quiet anti-glitches suggests that their potential origin may be
in the neutron star's interior~\citep{Duncan, Ranea-Sandoval, Garcia, Kantor}.
This possibility is also supported by the recent discovery of an anti-glitch
in the rotation-powered pulsar
PSR B0540-69~\citep{Tuo}.

The {\it Fermi} Large Area Telescope (LAT), which is continuously observing
more than 290 gamma-ray pulsars~\citep{Fermi-LAT}, is a powerful tool for
studying glitches~\citep{Sokolova}. An illustrative example of its
capabilities is the pulsar PSR J1522-5735, which was uncovered with a blind search in the LAT data spanning three years. The follow-up timing analysis of
gamma-ray emissions from this pulsar revealed a glitch that resulted in a
rapid decrease in the rotation frequency resembling an
anti-glitch~\citep{Pletsch:2013}. In this paper, we continue the follow-up timing analysis of PSR J1522-5735 using over 15 years of Fermi-LAT data to search for glitch events. We detected eight glitch events, six of which
were categorised as anti-glitches. 

The paper is organised as follows. {\it Fermi}-LAT data selection and
preparation procedures are explained in Section~\ref{sec:data}. In
Section~\ref{sec:Htest}, we introduce a phase model of pulsar rotation and
identify glitches by using the H-test statistic. The timing method used to
precisely measure phase model parameters is detailed in
Section~\ref{sec:timing}. The resulting timing solution is presented in
Section~\ref{sec:res}. In Section~\ref{sec:flux}, we perform emission
variability analysis. Finally, a summary is presented in Section~\ref{sec:dis}.

\section{Data preparation}
\label{sec:data}
The paper is based on the publicly available {\it Fermi}-LAT gamma-ray data 
for the time period from 2008 August 4 (54683 MJD) to 2024 April
5 (60405 MJD), which was prepared with {\it Fermi Science Tools} package
version v11r5p3. For the analysis, we included SOURCE-class photons according
to the P8R3\_SOURCE\_V3 instrument response functions with the energies
above $100$~MeV, within a $5^\circ$ region around the PSR J1522-5735
position, a zenith angle $< 100^\circ$, and when the LAT’s rocking angle
was less than $52^\circ$.

To increase sensitivity to pulsations, we calculated the weight $w_i$ for
each photon (i.e. the probability that the photon was emitted
by the pulsar; \citep{Bickel}).
We constructed a spectral model of gamma-ray sources
in the region around the PSR J1522-5735 that includes
{\it Fermi}-LAT 4FGL sources~\citep{Ballet} in a $10^\circ$ radius circle
as well as galactic and isotropic diffuse emission components.
The model parameters were optimised with binned likelihood analysis by
the {\it gtlike} tool. The spectral parameters of the pulsar and all sources
within $5^\circ$ and the normalisations of the background models were free to
vary in the fit. With the best-fitting source model, we used
{\it gtsrcprob} to compute the photons' weights based on their reconstructed
energy and arrival direction. A total of $500000$ photons with the
highest weights were kept for the subsequent analysis.
This corresponds to a probability-weight threshold of $0.022$ and 
$\sum_i w_i \simeq 30 030$.

Finally, the photon arrival
times were converted to the barycentric frame using the {\it gtbary} tool.
These computations were performed with the PSR J1522-5735's sky position
constrained to high precision by~\cite{Fermi-LAT}.

\section{Glitch identification}
\label{sec:Htest}

The analysis in this and the following sections employs the photon arrival times
$t_i$ and weights $w_i$ by relating them to a certain rotational phase model.
The model is usually given by a Taylor series expansion in time around a
chosen reference epoch $t_0$,
\begin{equation}
  \label{phase}
    \Phi(t) = \Phi_0 + f(t-t_0) + \frac{\dot{f}}{2}(t-t_0)^2 + \ldots \;,
\end{equation}
where $f$ is the pulsar frequency and $\dot{f}$ denotes the frequency
derivative over time.

As shown below, the PSR~J1522-5735 exhibited sudden deviations from the
model~\eqref{phase} corresponding to glitch events. A glitch occurring
at time $t_g$ with a permanent change in pulsar frequency $\Delta f_p$,
a frequency time derivative $\Delta \dot{f}_p$, and with frequency increment
$\Delta f_d$ decaying exponentially on a timescale of $\tau_d$ causes a phase
offset at time $t > t_g$ of
\begin{equation}
  \label{dphase}
  \Delta\Phi(t) = 
  \Delta f_p(t-t_g)+\frac{\Delta\dot{f}_p}{2}(t-t_g)^2 + \ldots +
  \Delta f_d\tau_d
  \left (1 - \mathrm{e}^{ -\frac{t-t_g}{\tau_d}} \right ) \;.
\end{equation}
We note that the instant changes in frequency and frequency derivative due to a
glitch are $\Delta f = \Delta f_p + \Delta f_d$ and $\Delta \dot{f} =
\Delta \dot{f}_p - \Delta f_d/\tau_d$, respectively.  

\begin{figure}
  \centerline{\resizebox{0.94\hsize}{!}{\includegraphics{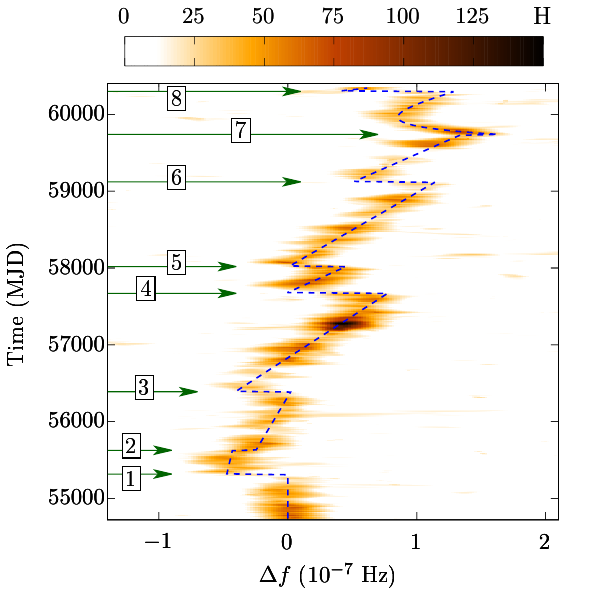}}}
  \caption{Weighted H-test for PSR J1522-5735
    calculated in 90\% overlapping 115-day data segments and
    with the phase model~\eqref{phase} at a fixed $\dot{f}$ and over a small
    range in $f$ centred on the pre-glitch value. The vertical axis shows
    the time midpoint of the data segment. The horizontal axis shows the
    offset in $f$ from the pre-glitch value. The weighted H-test is shown
    by the colour bar. The glitch events are noted by the arrows. The dashed
    blue curve is superimposed to show the timing solution given in
    Table~\ref{tab1}.}
   \label{fig1}
\end{figure}

The presence of glitches significantly complicates the timing analysis. In
order to identify a glitch in PSR J1522-5735 and evaluate the spin-parameter
changes associated with it, we applied the weighted H-test
statistic~\citep{Kerr}. We divided the data spanning more than 15 years into
115-day data segments with 90\% overlap. Then the value of $H$ was computed
separately for each segment~\citep{deJager} as

\begin{equation}
  \label{Htest}
  H = \max_{1 \leq L \le20} \left[
    \sum_{l=1}^{L}\mid \alpha_{l}\mid^{2}-4(L-1)
    \right]\;,
\end{equation}
where $\alpha_l$ is a Fourier amplitude of the $l$-th harmonic,  
\begin{equation}
  \label{alpha}
  \alpha_{l} = \frac{1}{\varkappa}\sum_{i}w_{i}\mathrm{e}^{-i l\Phi(t_i)}\;,
\end{equation}
with the normalisation constant 
\begin{equation*}
  \varkappa^2 = \frac{1}{2}\sum_{i}w_{i}^2\;.
\end{equation*}
The phase of each photon in Eq.~\eqref{alpha} was determined using a pre-glitch
phase model~\eqref{phase} in which we kept only the first two terms
of the Taylor series expansion according to the timing solution
of~\cite{Fermi-LAT}. We then fixed $\dot{f}$ and scanned a range in $f$ on a dense grid around a given
value. At each grid point, we computed $H$ using photons from the given data segment. The result is shown in 
Figure~\ref{fig1}. 

Visual inspection of the weighted H-test plot revealed sudden changes in
the pulsar's rotation rate around the epochs noted by the arrows in 
Figure~\ref{fig1}, which are candidates for glitches. From these results, we
estimated the values for the spin-parameter changes and glitch epochs, which
were refined by the timing procedure described in the following section.

\section{Timing analysis}
\label{sec:timing}

To accurately estimate the pulsar’s rotational and glitch parameters, we
used the timing procedure proposed by~\cite{Clark}, which is based on unbinned
likelihood maximisation. The rotational phase $\Phi$ is a
function of the photon arrival time $t_i$ and a set of phase model
parameters (see Eqs.~\eqref{phase},~\eqref{dphase}).
For a template pulse profile $F(\Phi)$, which is an analytic
approximation of a wrapped probability density function of the pulsar’s
rotational phase, the likelihood is
\begin{equation}
  \label{like}
        {\cal L}(\lambda) = \prod_i \left [
          w_iF(\Phi(t_i,\lambda),\lambda) + (1-w_i)
          \right]\;,
\end{equation}
where a set of phase models and template parameters are denoted by $\lambda$.  
The most likely values of $\lambda$ are obtained by maximisation of the
likelihood~\eqref{like}, which is unbinned in both phase (via the template
profile) and time. The advantage of this procedure is that there is no need
to construct a set of data segments for pulse times of arrival (TOAs)
determination.

The Bayesian information criterion~\citep[BIC;][]{Schwarz} was used to select
the model that best matches the data. It is defined as
\begin{equation}
  \mathrm{BIC} = -2\mathrm{log} ({\cal L}(\lambda))
  + k\,\mathrm{log} \left ( \sum_{i=1}^N w_i \right ) \;,
\end{equation}
where ${\cal L}$ is the maximum likelihood calculated at the best-fitting
parameters $\lambda$ and $k$ is the number of free parameters in the model.
The latter accounts for the dimensionality penalty to avoid overfitting.
When choosing between different models to describe the data, we favoured the one with a
minimum BIC.

\begin{table}
  \caption{\label{tab1} Inferred spin and glitch parameters of PSR
    J1522-5735.}

  \centering
  \begin{tabular}{l l}
    \hline\hline
    Parameter & Value \\
    \hline
    Right ascension, $\alpha$ (J2000.0) & $15^h22^m05.^s29$ \\
    Declination, $\delta$ (J2000.0) & $-57^\circ34^\prime58^{\prime\prime}.73$\\
    Epoch (MJD) & $57550$\\
    Weighted H-test & $2157$\\
    Spin frequency, $f$ (Hz) & $4.895137036_{-1\times10^{-9}}^{+3\times10^{-9}}$\\
    Frequency derivative, $\dot{f}$ ($10^{-12}$ Hz/s) & $-1.49673_{-1\times 10^{-5}}^{+1\times 10^{-5}}$
    \end{tabular}
  \centering
  \begin{tabular}{c c c c c c}
    \hline
  N & Epoch & $\Delta f_p$ & $\Delta \dot{f}_p$  & $\Delta f_d$ & $\tau_d$\\
  {} & MJD & $10^{-8}$\,Hz &  $10^{-16}$\,Hz/s & $10^{-8}$\,Hz & days\\
  \hline
  
  g1 & $55317_{-6}^{+3}$ & $-4.7_{-0.2}^{+0.4}$ & $2_{-3}^{+2}$ & - & - \\
  g2 & $55626_{-13}^{+18}$ & $1.8_{-0.3}^{+0.4}$ & $2_{-2}^{+3}$ & - & -\\
  g3 & $56389_{-5}^{+5}$ & $-4.3_{-0.1}^{+0.1}$ & $6.6_{-0.3}^{0.3}$ & - & -\\
  g4 & $57671_{-4}^{+3}$ & $-7.9_{-0.3}^{+0.3}$ & $5_{-2}^{+2}$ & - & -\\
  g5 & $58019_{-5}^{+4}$ & $-4.5_{-0.3}^{+0.3}$ & $-4_{-2}^{+2}$ & - & -\\
  g6 & $59122_{-4}^{+4}$ & $-6.4_{-0.2}^{+0.2}$ & $3.8_{-0.6}^{+0.6}$ & - & -\\
  g7 & $59740_{-25}^{+56}$ & $-12_{-9}^{+2}$ & $8_{-4}^{+15}$ & $15_{-4}^{+8}$&$130_{-23}^{+110}$\\
  g8 & $60299_{-4}^{+4}$ & $-9_{-2}^{+1}$ & $44_{-24}^{+26}$ & - & -\\
  \hline
  \end{tabular}
\tablefoot{The spin and glitch parameters correspond to the minimum of the
  BIC. Errors refer to the 68\% confidence interval of the posterior
  marginalised distributions. Terms that do not contribute to a lower BIC
  are excluded from the phase model; their parameter values are indicated
  by a dash.}
\end{table}

At the beginning, we constructed a template pulse profile based on the
segment of the data before the first glitch. With the pre-glitch spin parameters
of the pulsar, $f$ and $\dot{f}$, given by~\cite{Fermi-LAT},
the phase for each photon was calculated according to Eq.~\eqref{phase}.
The weighted pulse profile was obtained by binning the phase and summing the
photon weights in each phase bin. Based on these results, we introduced a
template pulse profile in the form of wrapped Gaussian peaks:
\begin{equation}
  \label{pulse}
  F(\Phi) = \left ( 1 - \sum_\alpha a_\alpha \right )
  + \sum_\alpha a_\alpha\,g(\Phi,\mu_\alpha,\sigma_\alpha).
\end{equation}
Here, $g(\Phi,\mu,\sigma)$ denotes a wrapped Gaussian peak
centred at phase $\mu$ with width $\sigma$:
\begin{equation}
  \label{gauss}
  g(\Phi,\mu,\sigma) = \frac{1}{\sigma \sqrt{2\pi}}
  \sum_{k=-\infty}^{\infty} \mathrm{exp} \left (
  -\frac{(\Phi + k -\mu)^2}{2 \sigma^2} \right ).
\end{equation}
The template was fitted to the weighted pulse profile by maximizing the
likelihood. We used the BIC to estimate the appropriate number of components to
include in the template, finding that two Gaussian peaks were sufficient
(see Figure~\ref{fig4}).

Together with the template pulse profile parameters, we then varied the
spin parameters to find the most likely values that maximise the likelihood. 
To explore the likelihood surface in the parameter space, we used the Markov
chain Monte Carlo sampling algorithm with parallel tempering, which has
demonstrated efficiency in multimodal sampling and optimisation problems. 
At the final stage, we verified that inclusion $\ddot{f}$ did not improve
the fit considerably according to the BIC.

The timing procedure outlined above was then used to refine the parameters of
the glitches. We consistently expanded the data by including time intervals
that covered the next glitches. The most likely glitch parameters were obtained
by maximising the likelihood. We added glitch parameters one by
one, performed Monte Carlo likelihood maximisation, and kept the new
parameter in the phase model if its inclusion led to a decrease in the BIC.
With the number of glitch parameters selected in this way, we performed a
final longer Monte Carlo run to obtain the value that maximises the
likelihood and uncertainty of each parameter.

\begin{figure}
  \centerline{\resizebox{0.94\hsize}{!}{\includegraphics{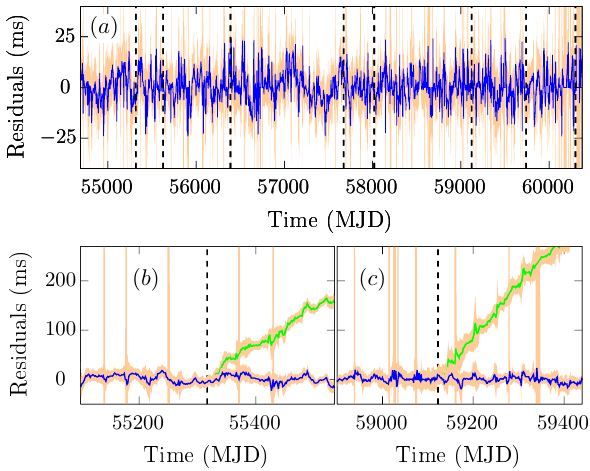}}}
  \caption{Measured timing residuals from timing solutions fitted by the unbinned
    likelihood maximisation method. (a) Residuals between the best-fit timing
    solution given in Table~\ref{tab1} over the data span. (b) The same as (a)
    but around the epoch of the first glitch event (blue line) in comparison
    with residuals between the pre-glitch model (green line). (c) The same as (a)
    but for the sixth glitch event. The shaded region corresponds to
    statistical uncertainties.}
   \label{fig2}
\end{figure}

\section{Results}
\label{sec:res}
The results of the timing analysis confirmed the eight glitch events pointed
out by the H-test. The minimum BIC timing solution is reported in
Table~\ref{tab1} and also noted by the dashed curve in Figure~\ref{fig1}.
It includes six anti-glitches (events g1, g3, g4, g5, g6, and g8 in
Table~\ref{tab1}) and two glitches (events g2 and g7 in Table~\ref{tab1}). Exponential post-glitch recovery terms for the anti-glitches and for the
glitch g2 were disfavoured by the BIC. For the glitch g7, the instant spin-up
with $\Delta f_p + \Delta f_d > 0$ (see Eq.~\eqref{dphase})
over-recovered to a net spin-down with $\Delta f_p < 0$.

We note that the inclusion of each glitch event into the timing model resulted
in a significant reduction in the BIC ($\Delta \mathrm{BIC} < - 30$). However,
glitch models incorporating decaying components and those without such
components yielded comparable BIC values. This indicates that the data cannot
reliably constrain the presence or absence of a post-glitch recovery phase.
Specifically, event g7 can alternatively be interpreted as an anti-glitch
without recovery occurring slightly later. This interpretation is supported
by the large uncertainties in the inferred glitch epoch and recovery timescale
$\tau_d$ for this event.

In order to demonstrate the validity of the phase model, we measured the
timing residuals --- the difference between the model-predicted and
the observed TOAs of pulsar pulses. The {\it Fermi}-LAT
data may contain only a few source photons superimposed on a strong background.
Consequently, a rigorous determination of TOAs necessitates the use of Poisson
statistics. For illustrative purposes, in this work the approach
of~\cite{Kerr:2015tva} was adopted. We computed the likelihood-based
cross-correlation of photon phases within 15-day-long segments with the
template pulse profile. Because the likelihood profile is asymmetric with
respect to phase change, we determined a phase interval around
its maximum bounded by a log-likelihood difference of
$|\Delta\, \mathrm{log}\, {\cal L}| = 2$. The centre of
this interval and half of its width was used to estimate the TOA and its
uncertainty.\footnote{For time segments with extremely low pulsar photon
  counts, the phase uncertainty interval expands to cover $\pi$, rendering
  the TOA undefined.}

The timing residuals are shown in Figure~\ref{fig2}. Their smallness
(Figure~\ref{fig2}(a)) clearly certifies the phase-coherence of the timing
solution. As one can see from the figure, the timing residuals exhibit
continuous low-level fluctuations. In the Monte Carlo analysis discussed
previously, we attempted to take these rotational irregularities into account
with the glitch model~\eqref{dphase}. However, this did not lead to a lower BIC
value. The residuals perturbed by the fluctuations are clearly distinguishable
from the PSR J1522-5735 glitches (e.g. see panels~(b) and (c) of
Figure~\ref{fig2}), but the presence of fluctuations can compromise the
accuracy with which the glitch recoveries can be studied.

Some pulsars have been observed to exhibit a spin-up glitch, where an
instantaneous increase in rotational frequency is followed by a rapid
over-recovery, resulting in a net spin-down
\citep[see, for example,][]{Archibald2017}. In sparse data, which fails to
resolve the initial spin-up and over-recovery processes, such a glitch
could be misclassified as an anti-glitch. This is a potential concern for
the {\it Fermi}-LAT, which observes a low photon count of
$\sum w_i \approx 5$ photons per day from PSR J1522-5735.

To assess whether the glitch events of PSR J1522-5735 are regular
spin-up glitches with an instant frequency change $\Delta f_p + \Delta f_d > 0$
followed by fast over-recovery to a net spin-down state, $\Delta f_p < 0$,
or anti-glitches with $\Delta f_p + \Delta f_d < 0$, a Bayesian
analysis was used.
For each detected glitch event, we prepared a data
segment that spanned from 115 days before the glitch began to 30 days before
the time of the next glitch. We then performed Monte Carlo runs within each
data segment to explore the parameter space of the glitch models, with the
decaying terms at a fixed template pulse profile and pre-glitch parameters.
Flat prior distributions for the free parameters within reasonably large
intervals were used in the runs.

For each data segment, we computed the Bayes factor for the models with
$\Delta f_p + \Delta f_d > 0$, which corresponds to a regular spin-up glitch
with a certain amount of over-recovery with respect to the anti-glitch models
with $\Delta f_p + \Delta f_d < 0$,
\begin{equation}
  B = \left .
  \int\limits_{\Delta f_p + \Delta f_d > 0} P(\lambda)\,d\lambda
  \;\; \right / 
  \int\limits_{\Delta f_p + \Delta f_d < 0} P(\lambda)\,d\lambda\;,
\end{equation}
where $P(\lambda)$ is the posterior probability distribution over free
parameters $\lambda$. The results are $0.066$ for the glitch events g1 and g8,
$0.12$ for g5 and greater than $0.18$ for the others. Thus, the possibility that the
glitch events g1, g5, and g8 observed in PSR J1522-5735 represent spin-up
glitches appears to be marginal.

\section{Pulse profile and energy flux}
\label{sec:flux}

\begin{figure}
  \centerline{\resizebox{0.94\hsize}{!}{\includegraphics{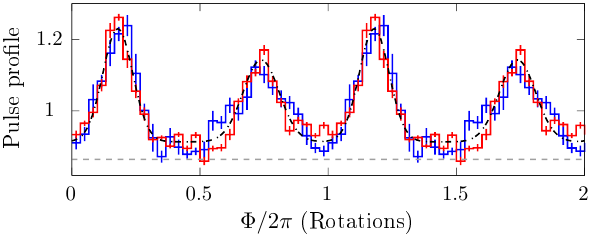}}}
  \caption{
    Weighted pulse profiles of PSR J1522-5735 averaged over Monte Carlo
    realisations for the timing model obtained using the data before
    the first (blue line) and after the third (red line) glitch events. The
    error bars show the standard deviation across the ensemble of
    realisations. The template pulse profile used for the timing analyses is
    shown with a black dashed-dotted line. The grey dashed line indicates the
    background level.
  }
  \label{fig4}
\end{figure}

In order to search for pulse profile changes induced by the glitches, we
performed Monte Carlo simulations similar to those previously described using
the same data segments but allowing the post-glitch pulse profile template to
vary during the runs. At each Monte Carlo step, we evaluated the weighted
pulse profile by `phase-folding' photon arrival times for a given realisation
of model parameters. Finally, the weighted pulse profile was averaged over
the realisations. We found no significant alteration in the pulse profile shape
following either the spin-up or the spin-down glitches (see Figure~\ref{fig4}).

\begin{figure}
  \centerline{\resizebox{0.94\hsize}{!}{\includegraphics{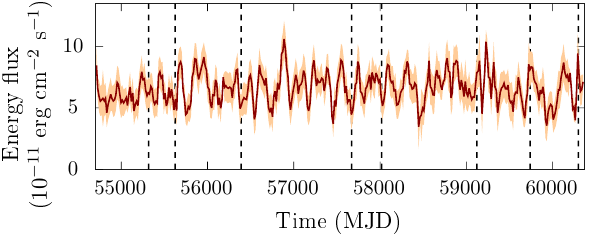}}}
  \caption{Energy flux from PSR J1522-5735 calculated in 75\% overlapping
    60-day data segments. The horizontal axis shows the time midpoint of the
    data segment. The shaded region corresponds to one-sigma uncertainties.}
   \label{fig3}
\end{figure}

It is also important to search for flux variations related to the glitches.
For this purpose, the gamma-ray flux from PSR J1522-5735 were monitored by
dividing the full data into 60-day-long data segments with 75\% overlap.
Based on the data from each segment, we constructed the spectral model of
the gamma-ray emission of the region around the PSR J1522-5735 as previously
described in Section~\ref{sec:data}. The spectral parameters of the
sources were fixed to the global ones due to the reduced exposure. The
spectral normalisation factors of PSR J1522-5735 and of the sources within
$5^\circ$ of the region around its position were set free and then optimised for every
time segment with binned likelihood analysis by the {\it gtlike} tool.
The energy flux calculated for each data segment is presented in
Figure~\ref{fig3}. As a result, we found no significant variations in the
flux observed around the epochs of the glitch events.

\section{Discussion}
\label{sec:dis}

We have carried out a comprehensive gamma-ray timing analysis of the pulsar PSR
J1522-5735 using the publicly available {\it Fermi}-LAT data spanning from 2008
August 4 to 2024 April 5. The analysis revealed eight glitch events during
this period: two spin-up glitches and six anti-glitches. The anti-glitches
are well fitted by a model with step-like changes in frequency and a
frequency time derivative with no post-glitch recovery terms. However, a
rapid recovery following anti-glitch within a few days cannot be reliably
constrained due to the sparsity of gamma-ray data. On the other hand,
  recoveries in a few days are shorter than those following the majority
  of glitches observed in rotation-powered pulsars, especially those
  characterised by a single exponential timescale.

The detection of six anti-glitches compared to two spin-up glitches in a
rotation-powered pulsar clearly has potential importance for the understanding of
these phenomena and necessitates a re-evaluation of existing models. Models
that explain anti-glitches in magnetars~\citep[see, for example,][]{Garcia}
are probably not applicable to the rotation-powered PSR J1522-3755, which does
not possess sufficiently strong magnetic fields. External models such as those
associated with strong outflows~\citep{Tong} or corona-mass-ejection-like
events~\citep{Lyutikov} can also explain the spin-down events. However, most
existing models predominantly yield to gradual deceleration processes, thereby
significantly constraining their capacity to generate sudden spin-down events.
Alternatively, \cite{Huang} propose that anti-glitches may arise
from the collision of a pulsar with a small celestial body. Such collisions
could induce a rapid spin-down of the pulsar's rotation rate. However, in
such a scenario, it is still difficult to explain the frequency of
anti-glitch events observed in PSR J1522-5735.

The observation of radiatively quiet anti-glitch events characterised by the
absence of significant variations in pulse profile shape and energy flux
suggests an internal origin for these phenomena. 
\cite{Kantor} proposed an anti-glitch model by extending the standard internal
scenario of~\cite{Anderson}. In this case, the glitch is due to momentum transfer between
the pulsar's superfluid and normal components in its inner crust or outer
core, depending on where the quantised vortices can pin. They proposed that
the superfluid fraction is dependent on the superfluid current and increases
as the velocity difference between the superfluid and normal components
decreases. In this case, under some condition, the pinned superfluid and the
rest of the star can decelerate, while the moment of inertia redistributes
through the formation of additional Cooper pairs to satisfy angular momentum
conservation, resulting in an anti-glitch. The higher the internal temperature
($\geq 10^7$ K) and the rotation frequency lag between the components
before vortex unpinning ($\geq 1$ rad/s), the more likely it is that an
anti-glitch occurs. PSR J1522-5735, with a characteristic age of $52$ kyr,
ranks among the youngest 4\% of known pulsars. Its youth implies elevated
internal temperatures and a dynamic rotational evolution, both of which favour
the development of a substantial rotational lag. These properties align with
the proposed prerequisites for the anti-glitch scenario of~\cite{Kantor}. 

While this study was in the final stage, the paper by~\cite{Zhou} appeared
with similar results. In that work, TOA fitting was used to ascertain glitch
parameters without employing BIC or alternative measures for model selection.
This may potentially clarify the differences in the glitch parameters
found in this study and in the referenced paper, as some exceed the two-sigma level.
Exploring these differences is beyond the scope of this paper and will be
addressed in future studies.

\begin{acknowledgements}
  We are indebted to M.Yu. Kuznetsov and G.I. Rubtsov for numerous
  inspiring discussions. The work is supported by the Russian Science Foundation
  grant 17-72-20291. The numerical part of the work is performed at the cluster
  of the Theoretical Division of INR RAS.
\end{acknowledgements}


\bibliographystyle{aa}
\bibliography{lit}
  
\end{document}